# Imaging and controlling electron transport inside a quantum ring

B. Hackens*, F. Martins*, T. Ouisse*, H. Sellier*, S. Bollaert†, X. Wallart†, A. Cappy†, J. Chevrier‡, V. Bayot§, S. Huant*

*Laboratoire de Spectrométrie Physique, Université Joseph Fourier Grenoble and CNRS, 140 rue de la Physique, 38402 Saint Martin d'Hères, FRANCE † IEMN, Villeneuve d'Ascq, FRANCE ‡ LEPES, CNRS, Grenoble, FRANCE § CERMIN, DICE Lab, UCL, Louvain-la-Neuve, BELGIUM

**Traditionally, the understanding of quantum transport, coherent and ballistic[1], relies on the measurement of macroscopic properties such as the conductance. While powerful when coupled to statistical theories, this approach cannot provide a detailed image of "how electrons behave down there". Ideally, understanding transport at the nanoscale would require tracking each electron *inside* the nano-device. Significant progress towards this goal was obtained by combining Scanning Probe Microscopy (SPM) with transport measurements[2-7]. Some studies even showed signatures of quantum transport in the surrounding of nanostructures[4-6]. Here, SPM is used to probe electron propagation *inside* an open quantum ring exhibiting the archetype of electron wave interference phenomena: the Aharonov-Bohm effect[8]. Conductance maps recorded while scanning the biased tip of a cryogenic atomic force microscope above the quantum ring show that the propagation of electrons, both coherent and ballistic, can be investigated *in situ*, and even be controlled by tuning the tip potential.**

An open quantum ring (QR) in the coherent regime of transport is a good example of interferometer : its conductance peaks when electron waves interfere constructively at the output contact and goes to a minimum for destructive interferences. Varying either the magnetic flux encircled by the QR or the electrostatic potential in one arm allows to tune the interference. This gives rise to the well-known magnetic[9] and electrostatic[10-11] Aharonov-Bohm (AB) oscillations. Although these effects have been studied extensively through transport measurements, those techniques lack the spatial resolution necessary to probe interferences in the interior of QRs. In this work, we perturb the propagation of electrons through a QR with an Atomic Force Microscope (AFM) tip. We therefore take advantage of both the imaging capabilities of the AFM and the high sensitivity of the conductance measurement to electron phase changes.

A 3D image of the QR used in the present work, as measured by our AFM in the conventional topographical mode, is shown in Fig. 1a. The QR is fabricated from an InGaAs/InAlAs heterostructure hosting a two-dimensional electron system (2DES) with a sheet density of $2 \times 10^{16}$ m$^{-2}$, buried 25 nm below the sample surface[12]. Electron-beam lithography and wet etching were used to pattern the QR and interconnections. At the experimental temperature (4.2 K), the QR is smaller than the intrinsic electron mean free path measured in the 2DES ($l_\mu = 2.3$ μm). Transport is thus in the ballistic regime with electrons travelling along "billiard-ball"-like trajectories. Moreover, the observation of periodic AB oscillations (inset of Fig. 1b) in the magnetoconductance of our QR attests that transport is also in the coherent regime[13]. The periodicity of these oscillations is found to be 26 mT, consistent with the average radius of circular electron trajectories in the QR: r = 220 nm.

The metallised tip of the AFM is biased at a voltage $V_{tip} = 0.3$ V and scanned in a plane parallel to the 2DES, at a typical tip-2DES distance of 50 nm. The tip acts as a flying nano-gate which modifies *locally* the electrostatic potential experienced by electrons within the QR and, hence, alters their transmission through the ring. The conductance of the sample, which is a measure of electron transmission, is recorded simultaneously in order to provide a 2D conductance map (Fig. 1c). The first-order contribution to this map is a broad background structure extending beyond the limits of the QR (note that a similar background was observed in the case of large open quantum dots[7]). Its overall shape is strongly affected by successive illuminations of the sample which are known to affect the configuration of ionized impurities. On the other hand, this background remains insensitive to B which is known to strongly affect quantum transport, both coherent and ballistic[12]. This indicates that the background is indeed related to a global shift of the electric potential in the *whole* quantum ring as the tip approaches the device.

A closer look to the conductance map reveals that the broad background is decorated by a more complex pattern of smaller-scale features, particularly visible in the central part of the image. We will see that this second-order effect shares a common feature with quantum transport inside the QR: its sensitivity to magnetic field. A high-pass filter applied to the raw conductance map reveals clear conductance oscillations (Fig. 1d) whose physical origin will be investigated in the remaining of the paper. The typical spatial periodicity of the oscillations (~100 nm) is much larger than the electron Fermi wavelength in our sample ($\lambda_F \sim 20$ nm). This tells us immediately that the "standing electron wave" pattern observed in previous experimental studies[5-6] is not the relevant mechanism to explain our observations. Note that the amplitude of the $\Delta G$ fringes on Fig. 1d is larger on the left side of the QR than on its right side. This left-right imbalance is observed whatever the direction of the magnetic field or the probe current. Therefore, it is related to an asymmetry of the QR (visible in Fig. 1a) or tip shape, and is not a signature of the Lorentz force which could also lead to an imbalance of the electron injection in the two arms of the QR. Most interestingly, data in Fig. 1d reveal that the type of fringe pattern depends on the scanned region. While fringes are predominantly *radial* when the tip is located directly above the ring, they become *concentric* when the tip moves away from the QR.

Fig. 2 shines light on the fundamental difference between both types of fringes as evidenced by the influence of an added bias current $I_{DC}$ on the conductance map. Increasing the absolute value of $I_{DC}$ is equivalent to raising the electron excess energy relative to the Fermi energy[14]. The bias current applied in Fig. 2b and d brings the electron system out of thermal equilibrium by more than 1 meV. We analyze the data by dividing the scan in three regions : the area enclosed in the QR (region I), a ring-shaped area in the vicinity of the QR (region II) and finally an area situated far from the QR where the influence of the tip vanishes (region III). First, we observe a strong reduction of concentric fringes (region II) at large bias disrespect of its sign. Region I, on the other hand, exhibits a very different behaviour depending on current direction: while conductance fringes die out at large positive bias (Fig. 2b), they strengthen and show a somewhat different pattern when bias is reversed (Fig. 2d).

A more quantitative picture of these differences is revealed in the evolution of $\delta G$, the standard deviation of $\Delta G$, calculated over regions I-III (Fig. 2a). While region III exhibits no dependence on $I_{DC}$ and hence serves as a reference for the noise level, region II shows a decrease of $\delta G$ at large bias, which is symmetric in $I_{DC}$. Since coherent effects are extremely sensitive to electron excess energy, which enhances the electron-electron scattering rates[14], this observation points towards a coherent origin for the concentric fringes in region II. By contrast, $\delta G$ in region I shifts gradually from one level to another as $I_{DC}$ is reversed. Keeping the magnetic field unchanged and reversing the current is equivalent to reversing the Lorentz force, which means that semi-classical ballistic trajectories inside the QR are rearranged. The behaviour observed in region I can hence be viewed as a sign of this rearrangement : when the tip scans above one arm of the QR, it changes the potential landscape, and this mainly affects the transmission of the semi-classical electron trajectories. While electrons are both coherent and ballistic in the QR, measurements in region I mainly reflects the changes of the pattern of ballistic semi-classical trajectories within the QR, and measurements in region II are predominantly determined by the coherent aspect of electron transport. As ballistic effects are only weakly sensitive to electron-electron scattering, this also explains the absence of a bell-shaped contribution to the δG curve for region I in Fig. 2a.

The magnetic field, as it tunes the phase of interfering electrons, brings further valuable information on the origin of the fringes. Figs. 3a-c show conductance maps measured at B=1.5T, 1.513T and 1.526T, covering a complete Aharonov-Bohm cycle, i.e. the magnetic flux encircled in the area of the QR changes by one flux quantum ($\phi_0$), implying that the phases of electron waves propagating through the two arms of the QR shift by $2\pi$. As shown in the supplementary video, concentric fringes expand continuously upon increasing B. This is more clearly shown in Figs. 3d-e that present the evolution of average conductance profiles taken along two radius of the QR (regions $\alpha$ and $\beta$ in Fig. 3a) over an Aharonov-Bohm cycle. On the left of the QR (region $\alpha$, Fig. 3d), the oscillation pattern smoothly shifts leftwards by *one* period as the Aharonov-Bohm phase - $\Delta\phi$ - increases by *one* flux quantum. Symmetrically, on the right of the QR (region $\beta$, Fig. 3e), the oscillation pattern moves rightwards by one period as $\Delta\phi$ moves from 0 to $\phi_0$. We interpret this behaviour as a scanning-gate induced electrostatic Aharonov-Bohm effect. Indeed, as the tip approaches the QR, either from left or right, the electrical potential mainly raises on the corresponding side of the QR. This induces a phase difference between electron wavefunctions travelling through the two arms of the ring, which causes the observed oscillations. Since the magnetic field applies an additional phase shift to electron wavefunctions, the V-shaped pattern formed by leftwards- and rightwards-moving fringes on Figs. 3d-e corresponds to iso-phase lines for the electrons. This observation ensures that our data are directly related to the Aharonov-Bohm effect and that concentric fringes originate from an interference effect of coherent electrons. While the observed oscillations are reminiscent of those reported in QRs with biased side gates[10-11], our experiment brings direct spatial information on interference effects, as we take advantage of the possibility to scan a localized perturbation across the sample.

We finally turn to the effect of the magnetic field on the central part of our conductance maps. As shown in Figs. 3a-c and in the supplementary video, the evolution of the central pattern with B, while complex, always remains slow and smooth. This strongly contrasts with the much more unpredictable and faster evolution observed in experiments performed on large open quantum dots[7]. We further note that the pattern of fringes observed at B=2T (Fig. 1d) is surprisingly similar to that visible at B = 1.5 T (Figs. 3a-c). This is a sign of the regular behaviour of the electron motion in QR, in opposition to the chaotic evolution of the electron dynamics characterizing large open quantum dots. However, since the Fermi wavelength is much smaller than the width of the QR arms, a complete description of the evolution of the central pattern with B would require in-depth simulation of the density of states inside the device taking into account the effect of the tip potential.

The present combination of AFM with transport measurement is very powerful for investigating electron interferences in *real space* and imaging ballistic transport at the local scale *inside* buried mesoscopic devices. One can also envision to use the technique to test electronic devices based on real-space manipulation of electron interferences, such as the electronic analogs of optical or plasmonic components[15] (resonators, Y-splitters, ...). *In situ* control over the electron potential would also provide a mean to design new ballistic devices with desired characteristics (beam splitters, multiterminal devices...). Therefore, our study paves the way for a wealth of experiments probing and controlling the local behaviour of charge carriers inside a large variety of open nano-systems.

**Methods**

**Heterostructure and 2DES parameters.** Our InGaAs heterostructure was grown by molecular beam epitaxy. The layer sequence of the heterostructure is as follows : InP substrate, 400 nm InAlAs buffer, 15 nm $In_{0.7}Ga_{0.3}As$ channel, 10 nm InAlAs spacer, a δ-doping plane (4 $10^{12}$ Si $cm^{-2}$) and 15 nm InAlAs barrier. The 2DES is located at the interface between the channel and spacer layers, i.e. 25 nm below the surface. Measurements on the unpatterned 2DES in the dark yield 100 000 $cm^2$/Vs for the electron mobility at 4.2 K.

**Scanning Gate Microscopy (SGM) technique.** The experiments are performed with a home-made cryogenic AFM, with commercial Pt/Ir-coated Si cantilevers. The movements of the tip in AFM topography mode are detected using a Fabry-Perot optical cavity (note that light is switched off in SGM mode). The tip bias $V_{tip}$ was adjusted in order to minimize the variations of the QR conductance during the scan, so that the system is in the small-perturbation regime. We found that $V_{tip}$ = 0.3 V fulfilled this condition. This value is consistent with a surface potential of ~ 0.3 V measured by Kelvin Probe Microscopy on an unpatterned heterostructure region next to the QR.

**Data filtering.** We use the following procedure to choose the cut-off frequency ($f_{cut}$) of the filter that we apply to our raw conductance maps. First, as we change $V_{tip}$, we notice that the background of the conductance map remains essentially unchanged, while the position of the small-range features is significantly affected. Therefore, after averaging over a sufficient number of conductance maps acquired with different $V_{tip}$, the small-range features are averaged away, leaving only the background. We then compare the Fourier transform of the averaged image with a typical Fourier transform of a raw conductance map. We can clearly distinguish between two frequency ranges. The low-frequency range, where the two Fourier transforms are essentially identical, corresponds to the broad background structure in the real-space conductance map. The high-frequency range, where the spectral content is much reduced in the averaged Fourier transform, is related to the small-range features that we want to isolate. The cut-off frequency $f_{cut}$ is chosen at the transition between both ranges, i.e. $f_{cut} = 4$ µm$^{-1}$.

**Supplementary Information** accompanies the paper on **www.nature.com/nature**.

The authors acknowledge fruitful discussions with M. G. Pala, and technical support from J.-F. Motte. The experimental setup was built thanks to the initial input of M. Stark. This work was supported by grants from the European Commission (Marie Curie EIF to B. H.), the FCT - Portugal (to F. M.) and the

CNRS (to V. B.). We acknowledge support from the "Action Concertée Nanosciences", French Ministry for Education and Research.

Correspondence and requests for materials should be addressed to B. H. (e-mail: bhackens@spectro.ujf-grenoble.fr) or S. H. (e-mail: serge.huant@ujf-grenoble.fr).

**Figure 1** Experimental setup and conductance maps **a** Three-dimensionnal reconstruction of the QR topography, based on an AFM image. The inner and outer ring diameters are 210 nm and 600 nm, respectively. Conductance maps are obtained by biasing the AFM tip and scanning it at a distance from the 2DES, as schematically represented, and simultaneously measuring the conductance of the QR. **b** Conductance of the QR as a function of the magnetic field, measured at 4.2 K using a lock-in technique, with a probe current of 20 nA and a frequency of 1.574 kHz. From the average conductance level, we deduce that the number of quantum modes in the openings of the QR is larger than 6. The inset shows the magnetoconductance over a small range of B, exhibiting the periodic h/e Aharonov-Bohm oscillations. Above B = 2.2T, as the cyclotron radius shrinks below the width of the QR arms and openings, electrons move along the edges of the device and Shubnikov-de Haas oscillations develop. **c** Conductance of the QR as a function of the tip position, with $V_{tip}$ = 0.3 V, $d_{tip}$ = 50 nm and B = 2T. A strong long-range background is clearly visible, with smaller-range fringes superimposed. The position of the QR, recorded in AFM topography mode before the conductance imaging, is indicated as a plain line. **d** High-pass-filtered conductance map, exhibiting small-range fringes.

**Figure 2** Evidence for a different behaviour of the standard deviation of radial and concentric fringes **a** δG (standard deviation of ΔG) as a function of the bias current $I_{DC}$, calculated in different areas (I-III) of the filtered conductance maps, indicated in Fig. 2b. The asymmetric evolution in region I is a signature of ballistic effects, influenced by the sign of the probe current, while the symmetric decay in region II is related to the coherent nature of electron waves. Drawn lines are guide to the eye. **b-c** filtered conductance maps recorded at B=5.95T, for different bias currents $I_{DC}$ : +0.8 µA (**b**), 0.0 µA (**c**) and -0.8 µA (**d**).

**Figure 3** Evolution of filtered conductance maps along an Aharonov-Bohm cycle **a-c** Filtered conductance maps recorded at increasing magnetic fields between B=1.5 T and 1.526T (over one Aharonov-Bohm cycle), showing a continuous and mostly periodic evolution of the pattern of fringes. In the vicinity of the ring, the fringe pattern undergoes a cyclic evolution over the sequence: starting at the position of a conductance maximum at the beginning of the cycle, one observes a minimum at the middle of the cycle, and a conductance maximum again at the end of the cycle. **d-e** average of horizontal conductance profiles in region $\alpha$ and $\beta$ (on both sides of the QR, defined by the dashed rectangles on Fig. 3a), plotted as a function of the phase difference $\Delta\phi$, in units of the Aharonov-Bohm period $\phi_0$. Note that on Figs. 3d-e, the overall lateral movement over a cycle is slightly larger for pattern $\alpha$ than for pattern $\beta$. This can be explained by a slight leftwards movement of the QR during the cycle, confirmed by a comparison of topographical images QR recorded before and after the sequence shown in Fig. 3a-c (the error bar over Fig. 3d represents the maximum extend of this movement). Note also that the color scale in Figs. 3d-e is smaller than on Figs. 3a-c and on the supplementary video, in order to evidence more clearly the movement of concentric fringes on the right of the QR. The evolutions of the patterns are opposite in the two regions (leftwards on Fig. 3d, and rightwards on Fig. 3e), and the V-shaped patterns correspond to iso-phase lines for the electron waves in the QR.

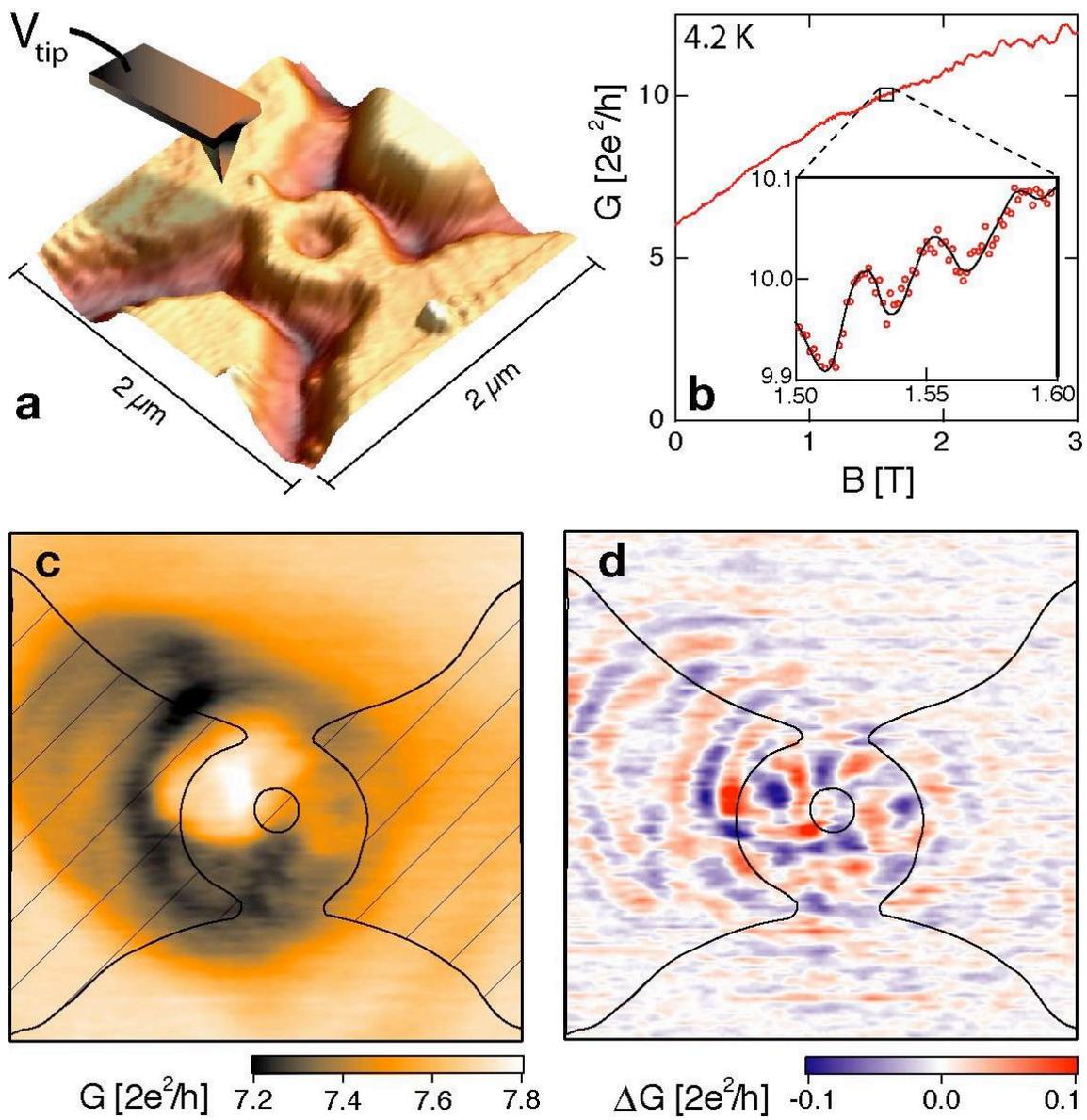

Fig. 1

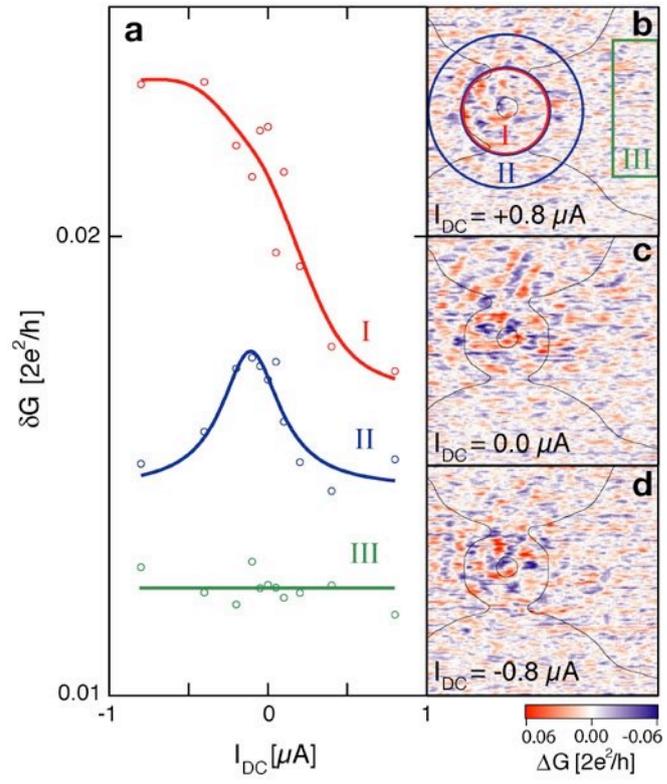

Fig. 2

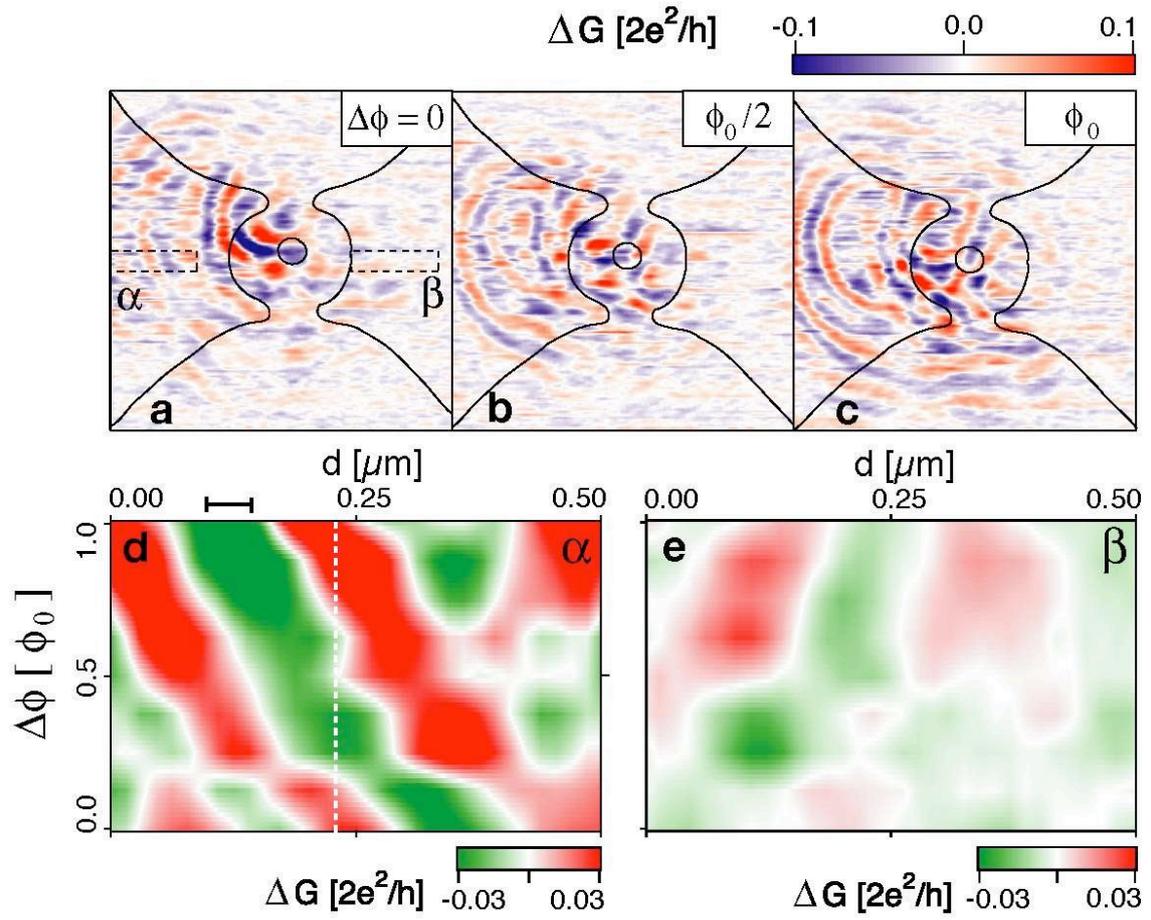

Fig. 3